\documentclass[
    ,final            
  ]
  {aipproc}

\layoutstyle{6x9}

\begin{document}

\title{On the $\eta_b \to J/\psi\ J/\psi$ decay}


\author{Pietro Santorelli}{
  address={Dipartimento di Scienze Fisiche, Universit{\`a} di
Napoli ``Federico II'', Italy\\
Istituto Nazionale di Fisica Nucleare, Sezione di Napoli, Italy
}}
\begin{abstract}
It has been argued long ago that $\eta_b$ could be observed through
the $\eta_b\to J/\psi(\to \mu^+\mu^-) J/\psi(\to \mu^+\mu^-)$ decay chain.
Recent calculations indicate that the width of $\eta_b$ into two $J/\psi$
is almost three order of magnitude smaller than the one into the $D \overline{ D^\ast} $.
We study the effects of final state interactions due to the
$D \overline{ D^{\ast}} $ intermediate state on the  $J/\psi\ J/\psi$ final state.
We find that the inclusion of this contribution may enhance the short
distance branching ratio of about two orders of magnitude.
\end{abstract}

\maketitle

About thirty years after the discovery of the $\Upsilon (1S)$ \cite{Upsilon1S},
no pseudoscalar $b\overline b$ states have been discovered.  The experimental
search of $\eta_b$ has been done at CLEO \cite{CLEO}, LEP \cite{ALEPH,L3,DELPHI} and CDF
by using different decay processes. In the following we will focus our attention on
the $\eta_b \to J/\psi\ J/\psi$ decay process to discover $\eta_b$ and we
will report the results obtained in \cite{Santorelli:2007xg}.

Six years ago the authors of ref. \cite{Braaten:2000cm} encouraged by
the large observed width of $\eta_c\to\phi\phi$ suggested to observe
$\eta_b$ through the $\eta_b \to J/\psi\ J/\psi$ decay process.
By using the measured branching ratio of $\eta_c \to \phi\phi$
and scaling laws with heavy quark masses the authors of ref.
\cite{Braaten:2000cm} obtained
\begin{eqnarray}\label{e:braaten}
{\cal B}r[\eta_b\to J/\psi\ J/\psi] & = & 7\times 10^{-4\pm 1}
\nonumber\\
{\cal B}r[\eta_b\to (J/\psi\ J/\psi)\to 4 \mu] & = & 2.5 \times 10^{-6\pm 1}.
\end{eqnarray}
Following this suggestion, CDF Collaboration has searched for
the $\eta_b \to J/\psi\ J/\psi\to 4\mu$ events in the full Run I data sample
\cite{Tseng:2003md}. In the search window, where a background of 1.8 events
is expected, a set of seven events are seen. This result seems confirm the
predictions in eq.~(\ref{e:braaten}).\\
Recently, Maltoni and Polosa \cite{Maltoni:2004hv} criticize the scaling procedure
adopted in ref. \cite{Braaten:2000cm} whose validity should reside only in the
domain of perturbative QCD. The non perturbative effects, which are dominant in
$\eta_c\to \phi\phi$, as a consequence of its large branching fraction, cannot be
rescaled by the same factor of the perturbative ones. In \cite{Maltoni:2004hv},
to obtain an upper limit on $\mathcal{B}r[\eta_b \to J/\psi\ J/\psi]$, the authors
evaluated the inclusive decay rate of $\eta_b$ to 4-charm states obtaining
\begin{equation}
{\cal B}r [\eta_b\to c\overline{c}c\overline{c}]=
1.8^{+2.3}_{-0.8}\times 10^{-5}\,,
\label{e:polosa}
\end{equation}
which is even smaller than the lower limit on
${\cal B}r[\eta_b\to J/\psi\ J/\psi]$ estimated in ref. \cite{Braaten:2000cm}.

Very recently Jia \cite{Jia:2006rx} have performed an explicit calculation
of the same exclusive $\eta_b \to J/\psi\ J/\psi$ decay process in the
framework of color-singlet model
\begin{eqnarray}
{\cal B}r[\eta_b\to J/\psi\ J/\psi] &\sim & (0.5-6.6)\times 10^{-8}\,,
\label{e:JiaJJ}
\end{eqnarray}
which is three order of magnitude smaller than the inclusive result in \cite{Maltoni:2004hv}.
The result in eq. (\ref{e:JiaJJ}) indicates that the cluster reported by
CDF~\cite{Tseng:2003md} is extremely unlikely to be associated with
$\eta_b$.  Moreover, the potential of discovering
$\eta_b$ through this decay mode is hopeless even in Tevatron Run II.\\
Another interesting decay channel to observe $\eta_b$, $\eta_b\to D^{(\ast)} \overline{D^\ast}$,
has been proposed in \cite{Maltoni:2004hv} where the range
$10^{-3}<{\cal B}r [\eta_b\to D \overline{D^\ast}]<10^{-2}$
was predicted. Finally in ref. \cite{Jia:2006rx} by doing reasonable physical considerations the author
estimated
\begin{eqnarray}
{\cal B}r [\eta_b\to D \overline{D^\ast}]
&\sim &  10^{-5}\,,\nonumber\\
{\cal B}r [\eta_b\to D^\ast \overline{D^\ast}]
&\sim &  10^{-8}\,,
\end{eqnarray}
which are at odds with the ones obtained in \cite{Maltoni:2004hv}.

\begin{figure}[h]
  \includegraphics[width=7truecm]{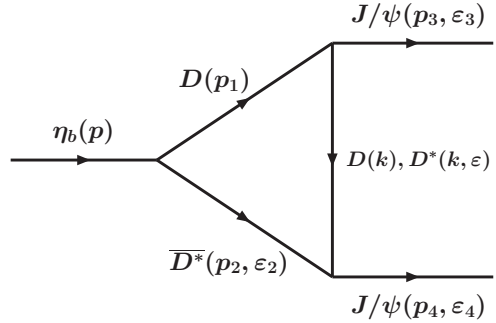}\\
  \caption{Long distance $t-$channel rescattering contributions to $\eta_b \to J/\psi\ J/\psi $.}
  \label{f:triangle}
\end{figure}

We study the $\eta_b\to J/\psi\ J/\psi$ by assuming that
\begin{itemize}
\item[a)] the branching ratio ${\cal B}r[\eta_b\to J/\psi\ J/\psi]$ is too small to be used to observe $\eta_b$ ( $\sim 10^{-8}$);
\item[b)] the branching ratio ${\cal B}r [\eta_b\to D \overline{D^\ast}] \sim  10^{-4\pm 1}$;
\item[c)] the ${\cal B}r [\eta_b\to D^\ast \overline{D^\ast}]$ is negligible in comparison with
${\cal B}r [\eta_b\to D \overline{D^\ast}]$,
\end{itemize}
and we will consider the effect of $ D \overline{D^\ast}\to J/\psi\ J/\psi$
rescattering (cfr figure \ref{f:triangle}). This  process should dominate the long distance
contribution to the decay under analysis due to the large coupling of $J/\psi$ to $ D^{(\ast)} D^\ast$ state (cfr later)
and the potentially large coupling of $\eta_b$ to  $D \overline{D^\ast}$ state \cite{Maltoni:2004hv}.

In particular we estimate the absorbitive part
of the diagram in fig. \ref{f:triangle} and we will neglect the dispersive
contribution. To evaluate the contribution we need the numerical values of the (on-shell) strong couplings
$g_{JD D}$, $g_{JDD^\ast}$ and $g_{JD^\ast D^\ast}$.%
\footnote{We use dimensionless strong couplings in all cases. In particular our $g_{JDD^\ast}/m_{J/\psi}$
corresponds to $g_{JDD^\ast}(GeV^{-1})$ more used in literature.} We take
the results coming from QCD Sum Rules \cite{QCDSR}, the Constituent Quark Meson model \cite{Deandrea:2003pv}
and a relativistic quark model \cite{RQM} which are
compatible each other: ($g_{JD D},g_{JDD^\ast},g_{JD^\ast D^\ast}$)= $(6, 12, 6)$.
To take into account the off-shellness of the exchanged $D^{(*)}$ mesons
in fig. \ref{f:triangle} we have introduced the $t-$dependance of the couplings
by means of the function
\begin{equation}
F(t)  = \frac{\Lambda^2 - m_{D^{(\ast)}}^2}{\Lambda^2-t}\,,
\end{equation}
which satisfy QCD counting rules.
In the numerical calculations the unknown parameter $\Lambda$ is allowed to vary
in the range $2.0 \div 2.4$ GeV. Moreover, we neglect the $D-D^\ast$ mass difference, $m_D = m_{D^\ast} \approx 1.9$ GeV.

\begin{figure}
  \includegraphics[width=8truecm]{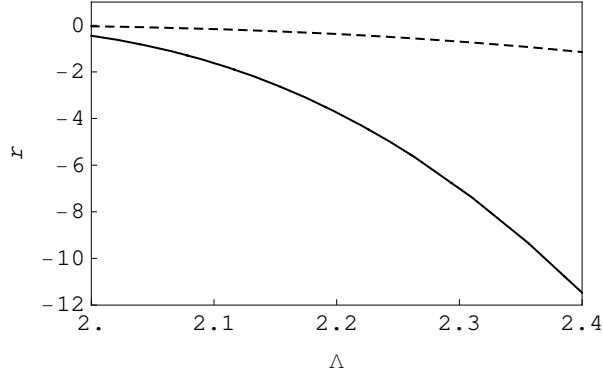}\\
  \caption{The  ratio $r$ (see text for definition) is plotted vs $\Lambda$ (in GeV) for $g_{\eta_b
DD^\ast}/g_{\eta_bJJ }\approx $ 4 (dashed line) and 40 (solid line).}\label{f:plot}
\end{figure}

The full amplitude for $\eta_b \to J/\psi\ J/\psi$ process can be written as
\begin{eqnarray}\label{e:fullA}
\!\!\!\!\!\!\mathcal{A}_f(\eta_b(p) \to J/\psi(p_3, \varepsilon_3)\ J/\psi(p_4, \varepsilon_4))\!\!\!\!\!\!\! & = &\!\!\!\!\!\!\!
\frac{\imath g_{\eta _b JJ }}{m_{\eta_b}}\, \varepsilon_{\alpha\beta\gamma\delta}
p_3^\alpha p_4^\beta\epsilon_3^{\ast \gamma} \epsilon_4^{\ast \delta}
\!\!\! \left[1+ \imath
\frac{g_{\eta _b DD^* }}{g_{\eta _b JJ }} A_{LD}\right]
\end{eqnarray}
where $A_{LD}$ represents the long-distance absorbitive contribution.
The effective couplings $g_{\eta_b DD^\ast}$  and $ g_{\eta_b JJ}$  defined by
\begin{eqnarray}
 \mathcal{A}(\eta_b(p)  \to  D(p_1)\ \overline{D^\ast}(p_2,\varepsilon_2)) & = &
 2\ g_{\eta_b DD^\ast}\;  (\varepsilon_2^\ast\cdot p)\,, \\
  \mathcal{A}(\eta_b(p)  \to  J/\psi(p_3, \varepsilon_3)\ J/\psi(p_4, \varepsilon_4)) & = &
   \frac{\imath g_{\eta_b JJ}}{m_{\eta_b}} \varepsilon_{\alpha\beta\gamma\delta}
p_3^\alpha p_4^\beta \varepsilon_3^{\ast\gamma}\varepsilon_4^{\ast\delta}\,,
\end{eqnarray}
and the ratio in eq. (\ref{e:fullA}) is obtained in terms of
theoretical estimation of the $\mathcal{B}r[\eta_b\to J/\psi\ J/\psi]/\mathcal{B}r[\eta_b \to D
\overline{D^\ast}] =10^{-4 \pm 1} $, \textit{i. e.} $g_{\eta_b
DD^\ast}/g_{\eta_bJJ } \approx 4 \div 40 $. This allow us to predict
the ratio between the long distance and short distance amplitude of the
$\eta_b \to J/\psi\ J/\psi$ decay process. In fig. \ref{f:plot}
the imaginary part of the amplitude, $r = A_{LD}\  g_{\eta _b DD^* }/g_{\eta _b JJ } $,
is plotted as a function of $\Lambda$ for the upper and lower bounds on the couplings ratio.
Looking at the figure we see that the long distance contribution coming from the graphs in
fig. \ref{f:triangle} is about ten times larger than the short distance amplitude.
It easy to show that, starting from the central value in eq. (\ref{e:JiaJJ}),
we predict the branching ratio in the range ($3.6\times  10^{-8} \div 5.0 \times 10^{-6})$,
where the lower bound corresponds to zero contribution from long distance and
the upper bound is obtained for $\Lambda = 2.4$ GeV and $g_{\eta_b DD^\ast}/g_{\eta_bJJ }\approx 40$.
Moreover, it should be observed that larger values of $\Lambda$ imply larger value of $r$.
On the other hand, $\Lambda$ should be not far from the physical mass of the
exchanged particle, the $D^{(\ast)}$ meson. To be specific, in an analysis on the
non-leptonic two body decays of B mesons taking into account final state
interaction with analogous graphs $\Lambda \approx 2.4$ GeV \cite{Cheng:2004ru}.
This value should represents an upper limit because of the larger mass in the
$s-$channel for $\eta_b$ decay.

As far as the number of events in full Tevatron Run I data (100 pb$^{-1}$) is concerned,
one should take into account the $\mathcal{B}r[J/\psi \to \mu^+\mu^-]\approx 6\%$
\cite{Yao:2006px} and the total cross section for $\eta_b$ production at Tevatron energy,
$\sigma_{tot}(\eta_b)=2.5 \ \mu b$ \cite{Maltoni:2004hv} obtaining between
0.03 and 5 produced $\eta_b$, where the range is due to the allowed range for
$\mathcal{B}r(\eta_b\to J/\psi\ J/\psi)$ .

This is compatible with the experimental data from CDF Collaboration on the Run I
dataset \cite{Tseng:2003md}. However, preliminary results from CDF Collaboration
Run II data at 1.1 fb$^{-1}$ \cite{paulini} seems to be at odds with the previous
findings. In fact, in the mass search window only 3 events has been observed.
We are looking forward for the publication of the final results for a comparison.

In conclusion, due to the large width of $\eta_b$ into $D \overline{D^\ast}$
final state, we have shown that the effects of final state interactions, \textit{i. e.} the
rescattering $D \overline{D^\ast} \to J/\psi\ J/\psi$, may increase
the $\mathcal{B}r[\eta_b \to 4 \mu]$ of about two orders of magnitude.
This result supports the experimental search of $\eta_b$ by looking at its decay
into $J/\psi\ J/\psi $, which has very clean signature.


\begin{theacknowledgments}
I would like to thank G. Nardulli for useful discussions.
\end{theacknowledgments}


\end{document}